\documentclass[draft]{agujournal2019}
\usepackage{url} 
\usepackage{lineno}
\usepackage{amsmath}
\usepackage[inline]{trackchanges} 
\usepackage{soul}

\usepackage{xcolor}

\draftfalse

\journalname{Geophysical Research Letters}

\begin{document}

%
%

\title{{A Systematic Look at the Temperature {Gradient} Contribution to the Dayside Magnetopause Current}}

%
%




\authors{J. M. H. Beedle\affil{1,2}, D. J. Gershman\affil{2}, V. M. Uritsky\affil{1,2}, T. D. Phan\affil{3}, B. L. Giles\affil{2}}


\affiliation{1}{Department of Physics, The Catholic University of America, Washington D.C. USA}
\affiliation{2}{NASA Goddard Space Flight Center, Greenbelt, MD, USA}
\affiliation{3}{Space Sciences Laboratory, University of California, Berkeley, CA, USA}




\correspondingauthor{Jason Beedle}{beedle@cua.edu}




\begin{keypoints}
\item {The magnetopause diamagnetic current is composed of opposing density and temperature gradient generated components}
\item  {The temperature gradient contributes up to {37\%} of the ion diamagnetic current density along the magnetopause}
\item The temperature component typically opposes the classical Chapman-Ferraro current direction

\end{keypoints}

%
%

%
%


\begin{abstract}

{Magnetopause diamagnetic currents arise from density and temperature driven pressure gradients across the boundary layer. While theoretically recognized, the temperature contributions to the magnetopause current system have not yet been systematically studied. To bridge this gap, we used a database of Magnetospheric Multiscale (MMS) magnetopause crossings to analyze diamagnetic current densities and their contributions across the dayside and flank magnetopause. Our results indicate that the ion temperature gradient component makes up to {37\%} of the ion diamagnetic current density along the magnetopause and typically opposes the classical Chapman-Ferraro current direction, interfering destructively with the density gradient component, thus lowering the total diamagnetic current density. This effect is most pronounced on the flank magnetopause. The electron diamagnetic current was found to be {5 to 14} times weaker than the ion diamagnetic current on average.}

\end{abstract}

\section*{Plain Language Summary}

The solar wind represents a continuous outflow of charged particles from the Sun’s upper atmosphere into the solar system. Upon reaching Earth's magnetosphere, the solar wind's dynamic pressure is balanced by the magnetic pressure of Earth's magnetic field in a boundary layer known as the magnetopause. This boundary layer represents the entry point of the solar wind's energy into Earth's magnetosphere and upper atmosphere, playing a crucial role in energy transport throughout the interconnected system. Plasma density and temperature differences across the boundary layer generate an electric current that supports the magnetopause. In this paper, we clarify the physical mechanism of the magnetopause current by using high-resolution data from NASA's MMS mission. We found a significant ion temperature contribution to the magnetopause current not identified in previous studies. Our results also indicated that the plasma electrons' contribution to the magnetopause current was significantly smaller than the ion contribution. 

%
%

%


%
%
%
%

\section{Introduction}

The magnetopause is a magnetosphere boundary layer created through the dynamic pressure balance between the solar wind's kinetic pressure and Earth's magnetic field. The solar wind causes distortions in the magnetosphere’s magnetic field topology supported by a current sheet first proposed by Chapman and Ferraro in 1931 \cite{CF1931}, often termed the Chapman-Ferraro (CF) current, which runs in a dawn-to-dusk direction around the magnetopause \cite{Ganushkina2017}. This current structure is believed to be generated through pressure gradients at the magnetopause boundary layer where, as explained in \citeA{Hasegawa2012}, the magnetosheath plasma has a higher plasma density, while the magnetosphere will have a higher ion temperature. The resulting changes in plasma density and temperature across the magnetopause leads to gradients that generate ion and electron diamagnetic currents running perpendicular to the magnetic field \cite{Ganushkina2017}. 

Because of the magnetopause's important role in magnetic reconnection and the resulting transfer of plasma and energy into the magnetosphere it has been the focus of numerous studies [\citeA{Cahill1963}; \citeA{Le1994}; \citeA{PhanLarson1996}; \citeA{PhanPaschmann1996}; \citeA{Haaland2014}; \citeA{Paschmann2018}; \citeA{Haaland2019}; \citeA{Shuster2019}; \citeA{Haaland2020}; etc.] and missions [MMS, THEMIS, and Cluster] which have delved deeper into the current sheet’s structure and creation. From \citeA{Paschmann2018} and their MMS magnetopause crossing database, the total current density across the dayside magnetopause was studied in detail. The flank magnetopause total current density was then surveyed in \citeA{Haaland2019} and \citeA{Haaland2020} where the flanks were found to have a weaker current density and a correspondingly thicker boundary layer than the dayside.

While the literature generally recognizes the importance of both the density and temperature in generating diamagnetic currents, a large scale systematic analysis of {their contribution to the} magnetopause current system has not yet been accomplished. To help fill this gap in the literature, we considered four years of burst mode MMS mission data over the magnetopause crossing intervals provided by \citeA{Paschmann2018}'s MMS database. We measured both the density and temperature diamagnetic current components and created current accumulations of their
contributions. In doing so, we found that the temperature diamagnetic current component is a statistically significant factor to both the dayside and flank magnetopause current sheet by acting against the density component and thus reducing the total diamagnetic current density. 

\section{Data and Analysis}

\subsection{MMS Database}
For this study, we used four years of data from \citeA{Paschmann2018,Paschmann2020}'s and \citeA{Haaland2020}'s database of MMS current sheet crossings. MMS is a mission comprised of four separate spacecraft traveling in a tetrahedron pattern through the magnetopause \cite{MMS_paper}. This database catalogues MMS current sheet crossings based on Fast Plasma Investigation (FPI) \cite{FPI}, Fluxgate Magnetometer \cite{FGM}, and Hot Plasma Composition Analyzer (HPCA) measurements \cite{HPCA}. The magnetopause transit times are captured through an automated minimum variance analysis {(MVA)} of the magnetic field in boundary normal or LMN coordinates \cite{Paschmann2018,Paschmann2020}. The start and end times of the current sheet crossing were then assigned as covering 76\% of the primary magnetic field component's {($B_L$'s)} transition across the magnetopause boundary, {which represents the thickness of a tanh fit of the $B_L$ profile}. 

{The magnetopause velocities for each crossing are obtained through a de Hoffman-Teller analysis and stored by the database as the de Hoffman-Teller velocity ($V_{HT}$). We then took these velocities and dotted them with the normal direction in the LMN coordinate system to get the normal magnetopause boundary velocity ($V_n$) \cite{Sonnerup1998,Khrabrov1998a,Paschmann2018}. The magnetopause thicknesses are then determined by taking this normal velocity and multiplying it by the time span of the current sheet as determined by the MVA analysis.}

\begin{figure}[htbp]
    \centering
    \noindent\includegraphics[width=0.90\textwidth]{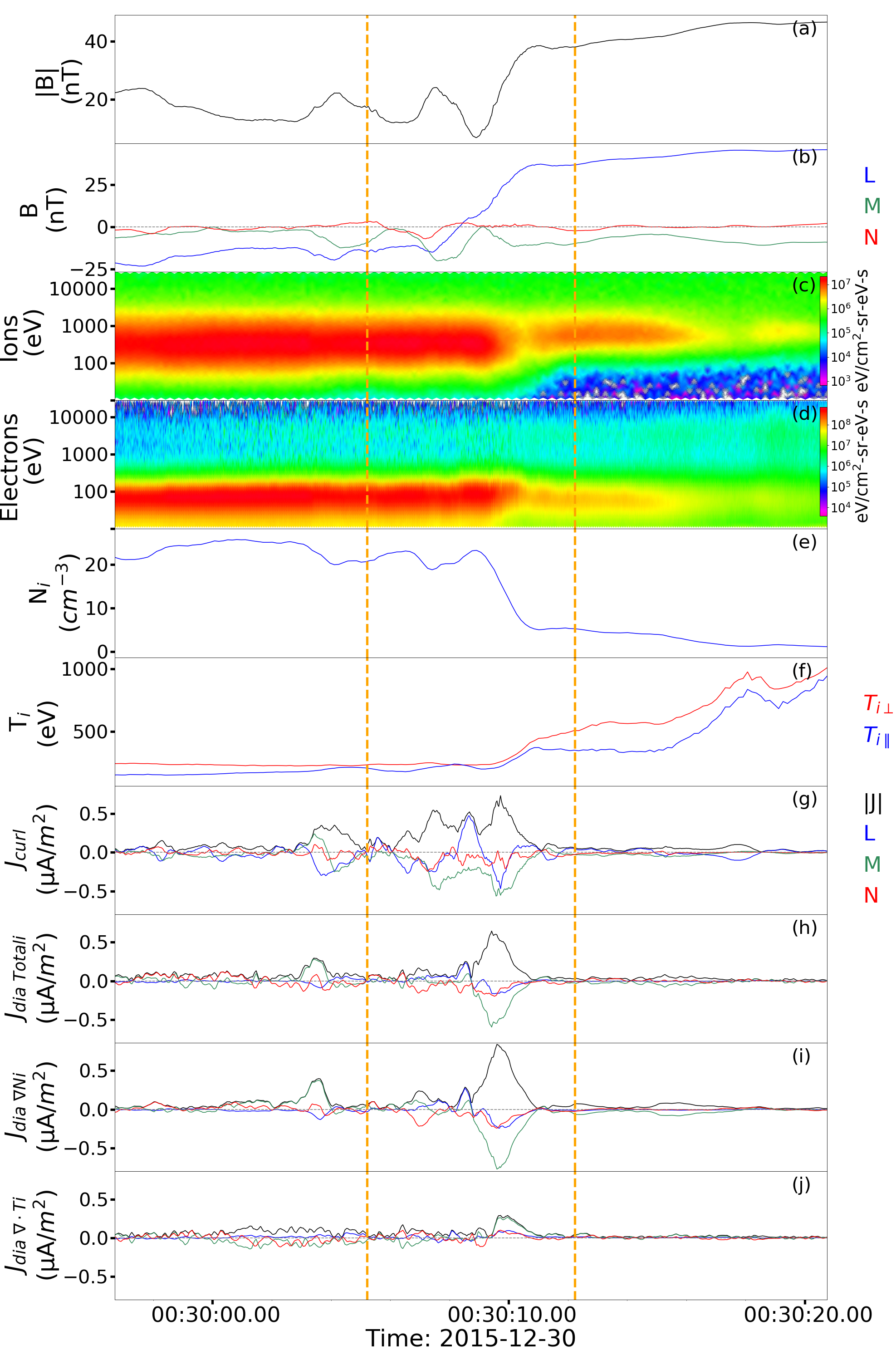}
    \caption{Example crossing during a monotonic (constant magnetopause velocity), complete 2015 MMS transit from the magnetosheath into the magnetosphere {at the following Cartesian GSE position along the dayside magnetopause, indicated in Earth Radii (X,Y,Z): (9.4 Re, -6.5 Re, -1.0 Re)}. The orange lines represent the magnetopause current sheet as identified by {the MMS database. (a) Magnitude of the magnetic field, (b) magnetic field in LMN coordinates, (c,d) ion and electron omnidirectional spectrograms, (e) ion number density, (f) ion perpendicular and parallel temperature, and (g,h,i,j) curlometer, total ion diamagnetic, density component, and temperature component current densities respectively in LMN coordinates with magnitudes indicated in black.}}
    \label{fig:example_crossing}
\end{figure}

The database places identifiers on the individually identified magnetopause crossings, classifying their characteristics and structure. A full description of this process and the current sheet identifiers can be found in \citeA{Paschmann2018} and \citeA{Haaland2020}. An example crossing, with the database defined magnetopause current sheet crossing indicated by the dashed orange lines, is given in Figure \ref{fig:example_crossing}. 

\subsection{Magnetopause Current Sheet Selection}
We chose database defined crossings from 2015 to 2018 using event identifiers to select for complete and monotonic magnetopause crossings, where monotonic indicated events that had a constant magnetopause velocity so their thicknesses and durations could be computed. Additionally, we included Harris sheet-like events, or simple clear magnetopause crossings that were also complete, monotonic events in our data set. Events with unusually high (above $2,000$ $nA/m^2$) current {density} spikes during the magnetopause crossing time, such as would occur during a reconnection event, or when data flags for the various instruments were non-nominal, were manually removed from our data set.

{Alongside the database defined event criteria, we imposed two additional conditions on our events in order to ensure high signal-to-noise ratios and typical magnetopause plasma number densities as described below:} 

{First, we considered the current measured during the magnetopause crossing and selected for events that reported significant peaks in their current densities. Where “significant" in our case was considered to be a current crossing where at least 50\% of the crossing duration was within 15\% of the maximum current peak during that crossing. This condition enabled us to select events with a strong current and high signal-to-noise ratios, thus ensuring that the temperature and density gradients and their resulting diamagnetic current components were not artificially diminished by the higher noise floor of low current density magnetopause crossings.}

{Second, we used HPCA measurements to compare the number densities of $H^{+}$ with $O^{+}$ for each magnetopause crossing. If $O^{+}$ exceeded $0.2$ $cm^{- 3}$ during the transit and $H^{+}$ remained below $1.5$ $cm^{- 3}$, then we considered $O^{+}$ to dominate the magnetospheric ion mass density by more than a factor of 2 as described by \citeA{Fuselier2019}. Events fitting this classification were also removed from our data set as they represented densities not typically found in the magnetopause current sheet.}

\begin{figure}[htbp]
    \centering
    \noindent\includegraphics[width=0.60\textwidth]{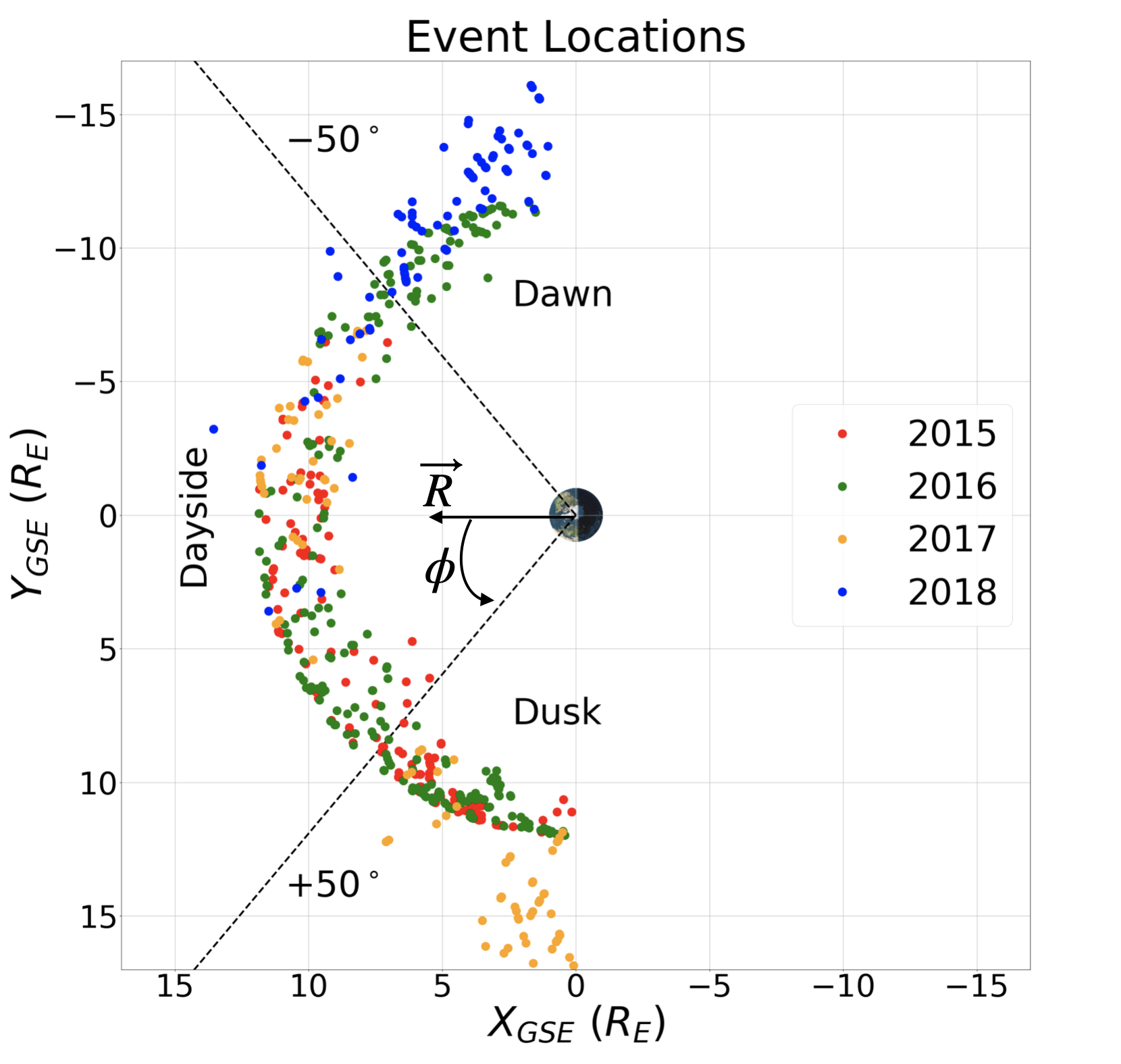}
    \caption{Diagram of our 561 MMS magnetopause crossings from 2015 (red), 2016 (green), 2017 (orange), and 2018 (blue). We define a local spherical coordinate system with $\phi$ in the $X_{GSE}$ - $Y_{GSE}$ plane, positively defined from the $+X_{GSE}$ axis, R defined as radially outward,  and $\theta$ as the {polar} angle into the $Z_{GSE}$ direction. Note, every $15^{\circ}$ in $\phi$ is equal to 1 hour of MLT with 12 MLT corresponding to $0^{\circ}$ in $\phi$, or along the +$X_{GSE}$ axis. {The Dawn flank is defined as $-50^{\circ}$ to $-90^{\circ}$ in  $\phi$, the Dusk flank as $+50^{\circ}$ to $+90^{\circ}$, and the Dayside as $+50^{\circ}$ to $-50^{\circ}$}. 
    Note, MMS first launched in 2015 with an orbit focusing on the dayside magnetopause, but after 2017, this orbit was extended to a wider orbit focusing on the flank magnetopause \cite{Haaland2020}. Because of the varying solar cycle, this has the possibility of creating an asymmetry between the dawn and dusk flank plasma measurements.}
    \label{fig:crossings}
\end{figure}

{From the application of the database identifiers as well as our conditions, we were able to identify 561 events. The locations and corresponding years of the selected crossings are denoted in Figure \ref{fig:crossings}.}

\subsection{Current Calculations}

MMS's four separate spacecraft allows the total current to be calculated by the curlometer method \cite{Dunlop1988} that uses all four spacecraft to perform the curl of the observed magnetic field in order to approximate Ampere's law in the MHD approximation {(e.g. \citeA{Ganushkina2017})}. This current we call the curlometer current: 

\begin{equation}
	\textbf{J}_{curl} = \frac{\nabla \times \textbf{B}}{\mu_0}.
\end{equation}

Using the curlometer method, we also calculated the gradient of the ion density, and the divergence of the temperature tensor to get the total ion diamagnetic current and its current components ($\textbf{J}_{dia \ \nabla N_i}$ and $\textbf{J}_{dia \ \nabla \cdot \overleftrightarrow{T_i}}$ ). We found the electron diamagnetic current to be at least one order of magnitude smaller, on average, than the ion diamagnetic current. This is in agreement with the results of \citeA{Dong2018} in their case study where they found that the ion diamagnetic current was the primary source of perpendicular current in the magnetopause. Thus we are presenting results for the ion components and will drop the signifier “i” going forward. The components {of the total diamagnetic current are found by imputing the ideal gas law, $\overleftrightarrow{P} = Nk_b \overleftrightarrow{T}$, into the expression for the total diamagnetic current $J_{\perp} = \frac{\vec{B} \times \nabla \cdot \overleftrightarrow{P}}{|B|^2}$, see e.g. \citeA{Ganushkina2017} and the references therein, resulting in the following expressions}:
 
\begin{equation}
	\textbf{J}_{dia \  \nabla N} = \frac{\textbf{B}\times(K_b\overleftrightarrow{T} \cdot \nabla N)}{|\textbf{B}|^2} \ \ , \ \  
	\textbf{J}_{dia \ \nabla \cdot \overleftrightarrow{T}} = \frac{\textbf{B}\times(K_bN\nabla \cdot \overleftrightarrow{T})}{|\textbf{B}|^2}.
\end{equation}

Where, by definition, $\textbf{J}_{dia \ Total} = \textbf{J}_{ dia \  \nabla N} + \textbf{J}_{dia \ \nabla \cdot \overleftrightarrow{T}}$.

{These currents were then averaged over our selected magnetopause crossings for each event, leading to single averaged current values for each of our 561 crossings. See Figure \ref{fig:example_crossing} for an example crossing on the dayside magnetopause. Note the two vertical orange lines represent the database defined current sheet crossing where the resulting currents would be averaged over.} 
 
All of the data taken from MMS, as well as the current calculations, was interpolated to the 30 ms FPI electron time resolution. As our main results involve ion diamagnetic currents and the total current as computed from the curlometer method, any sub 150 ms variations in the ion measurements will not impact our results. 
For non-curlometer calculations, we averaged over all four spacecraft to create a single data stream where, on average, the MMS separation during 2015 - 2018 was 10 - 60 km while the magnetopause current crossings analyzed typically had thicknesses greater than several hundred km, sufficiently larger than the max 60 km tetrahedron separation. These calculations were completed in GSE coordinates and then converted to a local spherical coordinate system built off of the Cartesian GSE coordinates. See Figure \ref{fig:crossings} for a depiction of our spherical coordinate system and the definition of the dusk and dawn sectors. 

{Note, one limitation with applying the curlometer method to a magnetopause current sheet crossing is that the curlometer method requires simultaneous measurements from all four spacecraft to calculate a gradient, curl, or divergence of a quantity. Thus, errors occur when one or more of the MMS constellation is outside of the magnetopause boundary as the spacecraft are no longer all measuring the same medium. For our study, this caveat is mitigated by the fact that the magnetopause boundary is, on average, at least one order of magnitude larger than the average MMS separation, with median magnetopause thicknesses often reported as approximating 1,000 km (e.g. \citeA{Haaland2019,Haaland2020}). This makes times where the curlometer method results in erroneous measurements brief and limited to the outskirts of a current sheet crossing.}

\subsection{Current Accumulations}

\begin{figure}[htbp]
    \centering
    \noindent\includegraphics[width=\textwidth]{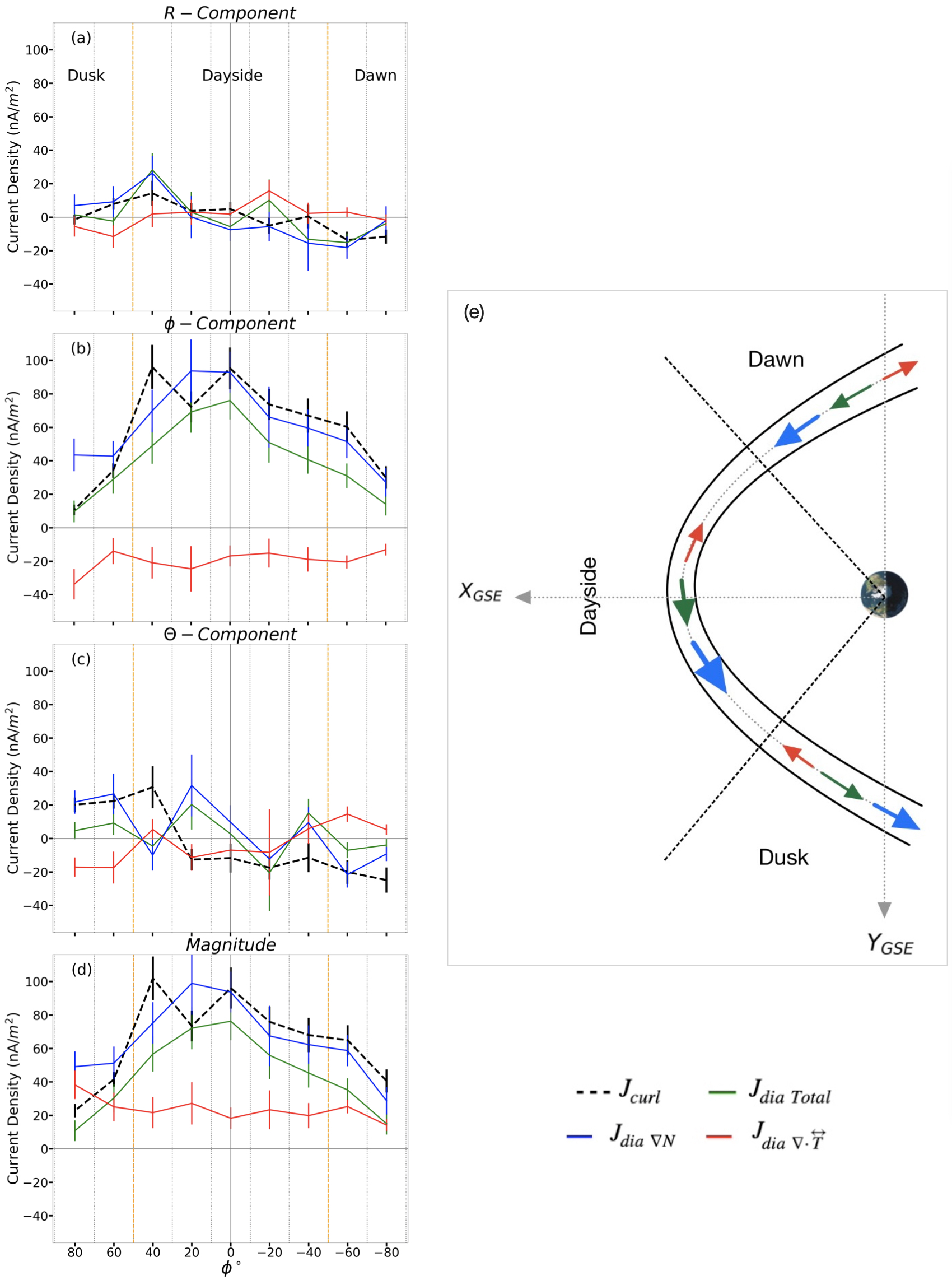}
    \caption{ a) through d) depict current density accumulations for $\textbf{J}_{curl}$, $\textbf{J}_{dia \ Total}$, $\textbf{J}_{dia \ \nabla N}$, and $\textbf{J}_{dia \ \nabla \cdot \overleftrightarrow{T}}$ across the dayside and flank magnetopause sectors, represented by the {orange} dashed lines. {The grey dashed lines represent the edges of the $20^{\circ}$ accumulation bins, centered at $80^{\circ}$, $60^{\circ}$, $40^{\circ}$, etc.} Moving from top to bottom: a). represents the R-component of the current in our local spherical coordinate system (described in Figure \ref{fig:crossings}). b). $\phi$ - component, c). $\theta$ - component, d). magnitude of the current components. e) diagram of the contributions and directions of $\textbf{J}_{dia \ Total}$, $\textbf{J}_{dia \ \nabla N}$, and $\textbf{J}_{dia \ \nabla \cdot \overleftrightarrow{T}}$ across the dusk, dayside, and dawn magnetopause. Note the size of the arrows in each sector indicates the relative magnitude of their current densities and the direction indicates the current component's flow around the magnetopause.}
    \label{fig:accumulations}
\end{figure}

The averaged currents from the 561 magnetopause crossings were put into bins corresponding to MMS’s physical location in our local spherical coordinate system. The angle $\phi$ was used to create 1-dimensional bins from $90^{\circ}$ to $-90^{\circ}$ in $20^{\circ}$ increments. This was done for each component in the spherical coordinate system as described in Figure \ref{fig:crossings} and shown in Figure \ref{fig:accumulations}. Error bars for each figure were computed using the standard error or $\sigma/\sqrt{N}$, with $\sigma$ the standard deviation of the values in each bin and N the number of events that fell inside that bin, {with the bins visually depicted in Figure 3 (a-d) by the grey-dashed lines.}

Using Figure \ref{fig:accumulations} we can make several observations. The first is that the $\textbf{J}_{curl}$, $\textbf{J}_{dia \ Total}$, and $\textbf{J}_{dia \ \nabla N}$ $\phi$-component currents are all in the $+ \phi$ direction across the magnetopause, or in the classical CF, dawn-to-dusk direction. However, the $\textbf{J}_{dia \ \nabla \cdot \overleftrightarrow{T}}$ $\phi$ component is consistently in the $-\phi$ direction, or from dusk-to-dawn across the magnetopause. Therefore the two components of the ion diamagnetic current, $\textbf{J}_{dia \ \nabla N}$ and $\textbf{J}_{dia \ \nabla \cdot \overleftrightarrow{T}}$, are oppositely directed across the magnetopause as can be seen in Figure \ref{fig:accumulations}b. The directions of these components are as expected when using the magnetospheric quantities evaluated by \citeA{Hasegawa2012} where the density component should run in the traditional CF current direction as the plasma density is higher in the magnetosheath and lower in the magnetosphere. At the same time, it is expected that the ion temperature is lower in the magnetosheath and higher in the magnetosphere, leading to the ion temperature component typically running counter to the CF current direction. On average, however, $\textbf{J}_{dia \ \nabla N}$ is stronger than $\textbf{J}_{dia \ \nabla \cdot \overleftrightarrow{T}}$, which allows the total diamagnetic current, $\textbf{J}_{dia \ Total}$, to still flow in the classical CF direction. 

\subsection{Current Measurement Results}

We then used our data to create a table of results over the dusk, dayside, and dawn magnetopause including the mean, median, and standard errors for our 561 magnetopause crossings as seen in Table \ref{tab:table1}.

{From this table, $\textbf{J}_{curl}$ is strongest on the dayside, with a dusk-dawn asymmetry as the dawn curlometer current is stronger than the dusk. Both $\textbf{J}_{dia \ Total}$ and $\textbf{J}_{dia \ \nabla N}$ show similar distributions with the dayside again being the strongest sector, but the dusk and dawn results are now in agreement within their standard errors. The $\textbf{J}_{dia \ \nabla \cdot \overleftrightarrow{T}}$ component shows a dusk-dawn asymmetry with a significantly stronger dusk current density than the dawn. The total electron diamagnetic current, $\textbf{J}_{e \ dia \ Total}$, is the weakest current component studied and shows a dusk-dawn asymmetry with the dusk being significantly weaker than both the dusk and the dayside.} 

{Comparing these results with past studies, our magnetopause thickness are, in general, higher than those found by \citeA{Haaland2020}; however, this is to be expected as we are considering a specific subset of database events as described in the previous sections. For similar reasons, the current densities found by \citeA{Haaland2020} show differences, with our dawn current densities matching closely, but the dusk and dayside values showing deviations.}

\newpage

\begin{table}[h]
    \caption{Magnetopause parameters, magnitudes, {and $\phi$ components} of current densities across Dusk: {($+50^{\circ}$ to $+90^{\circ}$), Dawn: ($-50^{\circ}$ to $-90^{\circ}$), and Dayside: ( $>$ $-50^{\circ}$ and $<$ $+50^{\circ}$)} crossings with the following format: mean (median) $\pm$ standard error of the values measured over each event's magnetopause crossing. Note, $V_n$ values are unsigned averages and medians of the normal magnetopause velocity.}
    \centering
    \begin{tabular}{c c c c}
    \hline
    Parameter & Dusk & Dayside & Dawn \\
    \hline
    Number of Crossings & 215 & 225 & 121  \\
    Thickness (km) & 1961 (1429) $\pm$ 132 & 2032 (1379) $\pm$ 135 & 1950 (1393) $\pm$ 180 \\
    Thickness ($d_i$) & 21.4 (14.3) $\pm$ 1.5 & 26.7 (16.4) $\pm$ 1.8 & 28.0 (15.9) $\pm$ 3.0 \\
    Thickness ($R_{gi}$) & 35.6 (18.7) $\pm$ 4.3 & 85.9 (33.2) $\pm$ 12 & 33.2 (17.9) $\pm$ 4.0 \\ 
    $V_n$ (km/s) & 131.5 (109.8) $\pm$ 7.0 & 99.8 (71.1) $\pm$ 7.5 & 94.3 (86.0) $\pm$ 6.3 \\ 
    Duration (s) & 16.0 (11.2) $\pm$ 0.8 & 25.0 (18.7) $\pm$ 1.3 & 20.6 (14.7) $\pm$ 1.3 \\
    \  \\
    $|\textbf{J}_{curl}|$ $(nA/m^2)$ & 30.0 (17.6) $\pm$ 3.2 & 83.2 (67.7) $\pm$ 5.6 & 52.1 (35.0) $\pm$ 6.3 \\
    $|\textbf{J}_{i \ dia \ Total}|$ $(nA/m^2)$ & 19.4 (8.4) $\pm$ 5.8 & 59.0 (47.5) $\pm$ 6.9 & 25.1 (17.5) $\pm$ 5.0 \\
    $|\textbf{J}_{dia \ \nabla N_i}|$ $(nA/m^2)$ & 50.0 (28.3) $\pm$ 8.5 & 78.2 (64.7) $\pm$ 7.2 & 43.6 (32.7) $\pm$ 6.6 \\
    $|\textbf{J}_{dia \ \nabla \cdot \overleftrightarrow{T}_i}|$ $(nA/m^2)$ & 31.5 (16.6) $\pm$ 7.0 & 20.0 (13.5) $\pm$ 5.8 & 19.6 (10.8) $\pm$ 2.9 \\
    \ \\ 
    $\textbf{J}_{curl} \ ({\phi})$ $(nA/m^2)$ & 21.1 (10.9) $\pm$ 2.7 & 83.0 (67.4) $\pm$ 5.3 & 45.3 (30.3) $\pm$ 5.9 \\
    $\textbf{J}_{i \ dia \ Total} \ ({\phi})$ $(nA/m^2)$ & 18.2 (8.3) $\pm$ 5.2 & 58.7 (47.3) $\pm$ 5.2 & 22.6 (16.6) $\pm$ 5.0 \\
    $\textbf{J}_{dia \ \nabla N_i} \ ({\phi})$ $(nA/m^2)$ & 43.2 (26.1) $\pm$ 6.7 & 78.0 (64.7) $\pm$ 6.7 & 39.4 (30.4) $\pm$ 6.6 \\
    $\textbf{J}_{dia \ \nabla \cdot \overleftrightarrow{T}_i} \ ({\phi})$ $(nA/m^2)$ & -25.0 (-16.1) $\pm$ 6.2 & -19.3 (-13.3) $\pm$ 4.1 & -16.8 (-10.7) $\pm$ 2.7 \\
    \ \\ 
    $|\textbf{J}_{e \ dia \ Total}|$ $(nA/m^2)$ & 1.7 (0.5) $\pm$ 0.7 & 5.9 (4.4) $\pm$ 1.0 & 5.4 (3.6) $\pm$ 0.7 \\
    \hline
    \end{tabular}
    \label{tab:table1}
\end{table}

\section{Temperature Gradient's Impact on the Magnetopause Current System}

Using our results from Figure \ref{fig:accumulations} and Table \ref{tab:table1}, we can posit three primary ways the ion temperature gradient impacts the magnetopause current {density} and, in doing so, create a 2D diagram to summarize our findings as shown in Figure \ref{fig:accumulations}e. 

\begin{enumerate}
  \item The divergence of the ion temperature tensor generates up to one third of the total ion diamagnetic current {density} in the $\phi$ direction.

    \ 
    
    Specifically, in the $\phi$ direction, {the ratio of the temperature-driven component to the density-driven component's contribution to the total ion diamagnetic current gives the following break down with} $\textbf{J}_{dia \ \nabla \cdot \overleftrightarrow{T}}$ making up {$37\%$} of the diamagnetic current {density} on the dusk, {$20\%$} on the dayside, and {$30\%$} on the dawn.
    
    \  
    
  \item $\textbf{J}_{dia \ \nabla \cdot \overleftrightarrow{T}}$ goes in the opposite direction of the classical Chapman-Ferraro Current.
  
    \  
    
    $\textbf{J}_{dia \ \nabla \cdot \overleftrightarrow{T}}$ is clearly in the $-\phi$ direction across the magnetopause when considering Figure \ref{fig:accumulations}b and Table 1. This results in $\textbf{J}_{dia \ \nabla \cdot \overleftrightarrow{T}}$ lowering the contribution of $\textbf{J}_{dia \ \nabla N}$, making the $\textbf{J}_{dia \ Total}$ less than $\textbf{J}_{curl}$ on average as seen in Table \ref{tab:table1}. 
    
    \ 
    
  \item $\textbf{J}_{dia \ \nabla \cdot \overleftrightarrow{T}}$’s contribution to the magnetopause current {density} becomes more important toward the flank magnetopause.
  
    \ 
    
    $\textbf{J}_{dia \ Total}$ and $\textbf{J}_{dia \ \nabla N}$ are strongest on the dayside and grow steadily weaker on the dusk and the dawn flanks, with both flanks showing similar results for the current densities. This is in contrast to $\textbf{J}_{dia \ \nabla \cdot \overleftrightarrow{T}}$ which {increases in strength, particularly on the dusk flank}, resulting in the total diamagnetic current being decreased even further by $\textbf{J}_{dia \ \nabla \cdot \overleftrightarrow{T}}$'s impact on the flank mangetopause.

\end{enumerate}

{From Table \ref{tab:table1} and Figure \ref{fig:accumulations} we may also notice the difference between the curlometer and the total diamagnetic current densities across the magnetopause. This difference is generally expected as the curlometer current represents the total current density as measured by MMS across the magnetopause current layer, which includes contributions from both ion and electron currents and their perpendicular and parallel components. The total ion diamagnetic current is thus one component of the curlometer current. This being said, the total diamagnetic current density still represents the main contributor to the curlometer current, accounting for almost 70\% of the current density on the dayside.}  

\section{Discussion}

\subsection{Gradient Formation}

These observations are generally consistent with previously literature regarding ion populations in the magnetosphere. From \citeA{Chappell2008}, the warm plasma cloak is defined as a population of 10 eV to 3 keV ions energized in the polar cap and magnetotail, which circulates in a dawn-to-dusk circulation pattern throughout the inner magnetosphere out to the magnetopause. As \citeA{Chappell2008} notes, the warm plasma cloak ions can occupy the same space as the much warmer and more energetic ring current ions, which circulate in the opposite direction across the magnetosphere, from dusk-to-dawn. Thus it is possible, on a simplified level of magnetospheric circulation, that the colder warm plasma cloak ions provide generating pressure for the density gradient component across the magnetopause in its dawn-to-dusk, CF current like direction, while the warmer ring current ions provide generating pressure for the temperature gradient component in its dusk-to-dawn direction. The density component's dayside-flank asymmetry could also be explained by additional density gradients generated by the plasmasphere drainage plume \cite{Borovsky2008}, which exhausts through the dayside magnetopause during storm conditions, enhancing the dayside with more cold ions, thus leading to an enhanced dayside $\textbf{J}_{dia \ \nabla N}$ while leaving the dusk and dawn components reliant solely on the warm plasma cloak ion population. {The presence of magnetosphere particle populations in the mangetopause and their effects on the magnetopause current sheet, specifically regarding magnetic reconnection, is explored in the following studies: e.g. \citeA{Borovsky2008}, \citeA{Fuselier2017}, and \citeA{Walsh2013}. Additional study regarding the effects of the warm plasma cloak and plasmasphere particle populations on the diamagnetic current and its} component generation is needed however.

\subsection{Large Scale Current System Closure}

{From Figure \ref{fig:accumulations} and Table \ref{tab:table1}, we can see the total diamagnetic current density is lower on the flanks, helped by an increasingly prominent temperature component. This indicates current closure with the larger 3D current system where the magnetopause current steadily curves toward the parallel as it goes around the flanks. As the total current through the magnetopause current system must remain constant, as charge is conserved, the current density of the total current's components must then fluctuate appropriately along the dayside and flanks to “transfer" the current density from the perpendicular dominated dayside to the increasingly parallel dominated flanks. This indicates that the total diamagnetic (perpendicular) current decreases in order for more of the total current density to be diverted toward parallel currents such as the field aligned currents. Additionally, the magnetopause is generally thinner on the dayside and thicker on the flanks while the magnetopause current must remain the same as it flows through the magnetopause (e.g. \citeA{Haaland2019,Haaland2020}). Thus the total perpendicular current density must change to compensate for either a thinner or thicker magnetopause. From Figure \ref{fig:accumulations}, we can see this is, indeed, the case as the total diamagnetic current density decreases on the flanks and is strongest toward the dayside.} 

\subsection{Ion vs. Electron Diamagnetic Current Densities}

Even though the electron diamagnetic current was found to be significant on electron scale current sheets by \citeA{Shuster2019,Shuster2021}, we found the electron current {density} to be less significant over our ion scale magnetopause current sheets. Specifically the electron diamagnetic current density is {$7\%$} of the ion diamagnetic current density in the $\phi$ direction on the dusk, {$10\%$} on the dayside, and {$20\%$} on the dawn. This presents an interesting asymmetry for the electron current as it is noticeably weaker on average on the dusk than it is on the dawn; however, in both cases, the electron diamagnetic current is also weaker than the contribution made by the ion current. The weaker electron current density may be explained based on the fact that we are averaging over many electron scale current sheets when considering our ion scale magnetopause crossing, thus lowering the resulting current density.

\section{Conclusions}

{In this paper, we have quantified the diamagnetic current's composite nature. Based on our systematic analysis of four years of MMS magnetopause crossings, the diamagnetic current is composed of two competing components: one generated by density gradients and one by temperature gradients.} 

{We have found that the temperature generated component acts against the density component, weakening the total diamagnetic current's net strength, particularly on the flanks where the temperature component's contribution can reach up to {37\%} of the diamagnetic current density along the magnetopause. We also found that ions contribute the majority of the current density to the diamagnetic current, with electrons accounting for only {7\%} to {20\%} of the ion's contribution.} 

{Taking these findings into account, we can posit that enhancements of the temperature gradient along the magnetopause boundary may lead to a corresponding weakening of both the diamagnetic current and, by extension, the total current in the magnetopause. This implied weakening of the magnetopause current by the temperature gradient leads to a more complicated picture of the interaction between the solar wind and Earth's magnetosphere, especially on small scales where situations can arise where the two components of the diamagnetic current become equal yet opposite, leading to the net cancellation of the diamagnetic current in that region. Studying the small-scale consequences of this interaction is the basis of our future work.}

\acknowledgments
The MMS current sheet database was created by Goetz Paschmann and Stein Haaland, and further developed by the International Space Science Institute Team 442, “Study of the physical processes in magnetopause and magnetosheath current sheets using a large MMS database". We thank them as well as the entire MMS team and instrument leads for the data access and support. We also thank the pySPEDAS team for their support and data analysis tools. This research was supported by the NASA Magnetospheric Multiscale Mission in association with NASA contract NNG04EB99C. J. M. H. B. and V. M. U. were supported through the cooperative agreements NNG11PL10A and 80NSSC21M0180. 

\section*{Open Research}
\noindent The MMS data used in this study is publicly available at https://lasp.colorado.edu/mms/sdc/public/datasets/ from the FPI, 
FIELDS, and HPCA datasets. The averaged MMS data over the 561 MP crossings used for this study are available through a Harvard Dataverse public database: https://doi.org/10.7910/DVN/SRBJCR. 



%
%

\bibliography{main.bib }

\begin{thebibliography}{}

\bibitem [\protect \citeauthoryear {%
Borovsky%
\ \BBA {} Denton%
}{%
Borovsky%
\ \BBA {} Denton%
}{%
{\protect \APACyear {2008}}%
}]{%
Borovsky2008}
\APACinsertmetastar {%
Borovsky2008}%
\begin{APACrefauthors}%
Borovsky, J\BPBI E.%
\BCBT {}\ \BBA {} Denton, M\BPBI H.%
\end{APACrefauthors}%
\unskip\
\newblock
\APACrefYearMonthDay{2008}{}{}.
\newblock
{\BBOQ}\APACrefatitle {A statistical look at plasmaspheric drainage plumes} {A
  statistical look at plasmaspheric drainage plumes}.{\BBCQ}
\newblock
\APACjournalVolNumPages{Journal of Geophysical Research: Space
  Physics}{113}{A9}{}.
\newblock
\begin{APACrefDOI} \doi{10.1029/2007JA012994} \end{APACrefDOI}
\PrintBackRefs{\CurrentBib}

\bibitem [\protect \citeauthoryear {%
Burch%
, Moore%
, Torbet%
\BCBL {}\ \BBA {} Giles%
}{%
Burch%
\ \protect \BOthers {.}}{%
{\protect \APACyear {2015}}%
}]{%
MMS_paper}
\APACinsertmetastar {%
MMS_paper}%
\begin{APACrefauthors}%
Burch, J\BPBI L.%
, Moore, T\BPBI E.%
, Torbet, R\BPBI B.%
\BCBL {}\ \BBA {} Giles, B\BPBI L.%
\end{APACrefauthors}%
\unskip\
\newblock
\APACrefYearMonthDay{2015}{}{}.
\newblock
{\BBOQ}\APACrefatitle {Magnetospheric Multiscale Overview and Science
  Objectives} {Magnetospheric multiscale overview and science
  objectives}.{\BBCQ}
\newblock
\APACjournalVolNumPages{Space Science Review}{199}{}{5-21}.
\newblock
\begin{APACrefDOI} \doi{10.1007/s11214-015-0164-9} \end{APACrefDOI}
\PrintBackRefs{\CurrentBib}

\bibitem [\protect \citeauthoryear {%
Cahill%
\ \BBA {} Amazeen%
}{%
Cahill%
\ \BBA {} Amazeen%
}{%
{\protect \APACyear {1963}}%
}]{%
Cahill1963}
\APACinsertmetastar {%
Cahill1963}%
\begin{APACrefauthors}%
Cahill, L\BPBI J.%
\BCBT {}\ \BBA {} Amazeen, P\BPBI G.%
\end{APACrefauthors}%
\unskip\
\newblock
\APACrefYearMonthDay{1963}{}{}.
\newblock
{\BBOQ}\APACrefatitle {The boundary of the geomagnetic field} {The boundary of
  the geomagnetic field}.{\BBCQ}
\newblock
\APACjournalVolNumPages{Journal of Geophysical Research}{68}{7}{1835-1843}.
\newblock
\begin{APACrefDOI} \doi{10.1029/JZ068i007p01835} \end{APACrefDOI}
\PrintBackRefs{\CurrentBib}

\bibitem [\protect \citeauthoryear {%
Chapman%
\ \BBA {} Ferraro%
}{%
Chapman%
\ \BBA {} Ferraro%
}{%
{\protect \APACyear {1931}}%
}]{%
CF1931}
\APACinsertmetastar {%
CF1931}%
\begin{APACrefauthors}%
Chapman, S.%
\BCBT {}\ \BBA {} Ferraro, V\BPBI C\BPBI A.%
\end{APACrefauthors}%
\unskip\
\newblock
\APACrefYearMonthDay{1931}{}{}.
\newblock
{\BBOQ}\APACrefatitle {A new theory of magnetic storms} {A new theory of
  magnetic storms}.{\BBCQ}
\newblock
\APACjournalVolNumPages{Terrestrial Magnetism and Atmospheric
  Electricity}{36}{}{77-97}.
\PrintBackRefs{\CurrentBib}

\bibitem [\protect \citeauthoryear {%
Chappell%
, Huddleston%
, Moore%
, Giles%
\BCBL {}\ \BBA {} Delcourt%
}{%
Chappell%
\ \protect \BOthers {.}}{%
{\protect \APACyear {2008}}%
}]{%
Chappell2008}
\APACinsertmetastar {%
Chappell2008}%
\begin{APACrefauthors}%
Chappell, C\BPBI R.%
, Huddleston, M\BPBI M.%
, Moore, T\BPBI E.%
, Giles, B\BPBI L.%
\BCBL {}\ \BBA {} Delcourt, D\BPBI C.%
\end{APACrefauthors}%
\unskip\
\newblock
\APACrefYearMonthDay{2008}{}{}.
\newblock
{\BBOQ}\APACrefatitle {Observations of the warm plasma cloak and an explanation
  of its formation in the magnetosphere} {Observations of the warm plasma cloak
  and an explanation of its formation in the magnetosphere}.{\BBCQ}
\newblock
\APACjournalVolNumPages{Journal of Geophysical Research}{113}{A09206}{}.
\newblock
\begin{APACrefDOI} \doi{10.1029/2007JA012945} \end{APACrefDOI}
\PrintBackRefs{\CurrentBib}

\bibitem [\protect \citeauthoryear {%
Dong%
\ \protect \BOthers {.}}{%
Dong%
\ \protect \BOthers {.}}{%
{\protect \APACyear {2018}}%
}]{%
Dong2018}
\APACinsertmetastar {%
Dong2018}%
\begin{APACrefauthors}%
Dong, X\BPBI C.%
, Dunlop, M\BPBI W.%
, Wang, T\BPBI Y.%
, Cao, J\BPBI B.%
, Trattner, K\BPBI J.%
, Bamford, R.%
\BDBL {}Torbert, R\BPBI B.%
\end{APACrefauthors}%
\unskip\
\newblock
\APACrefYearMonthDay{2018}{}{}.
\newblock
{\BBOQ}\APACrefatitle {Carriers and Sources of Magnetopause Current: MMS Case
  Study} {Carriers and sources of magnetopause current: Mms case study}.{\BBCQ}
\newblock
\APACjournalVolNumPages{Journal of Geophysical Research: Space
  Physics}{123}{}{5464--5475}.
\newblock
\begin{APACrefDOI} \doi{https:// doi.org/10.1029/2018JA025292} \end{APACrefDOI}
\PrintBackRefs{\CurrentBib}

\bibitem [\protect \citeauthoryear {%
Dunlop%
, Southwood%
, Glassmeier%
\BCBL {}\ \BBA {} Neubauer%
}{%
Dunlop%
\ \protect \BOthers {.}}{%
{\protect \APACyear {1988}}%
}]{%
Dunlop1988}
\APACinsertmetastar {%
Dunlop1988}%
\begin{APACrefauthors}%
Dunlop, M\BPBI W.%
, Southwood, D\BPBI J.%
, Glassmeier, K\BHBI H.%
\BCBL {}\ \BBA {} Neubauer, F\BPBI M.%
\end{APACrefauthors}%
\unskip\
\newblock
\APACrefYearMonthDay{1988}{}{}.
\newblock
{\BBOQ}\APACrefatitle {Analysis of multipoint magnetometer data} {Analysis of
  multipoint magnetometer data}.{\BBCQ}
\newblock
\APACjournalVolNumPages{Advanced Space Research}{8}{}{273-277}.
\PrintBackRefs{\CurrentBib}

\bibitem [\protect \citeauthoryear {%
Fuselier%
\ \protect \BOthers {.}}{%
Fuselier%
\ \protect \BOthers {.}}{%
{\protect \APACyear {2019}}%
}]{%
Fuselier2019}
\APACinsertmetastar {%
Fuselier2019}%
\begin{APACrefauthors}%
Fuselier, S\BPBI A.%
, Mukherjee, J.%
, M.~H.~Denton, S\BPBI M\BPBI P.%
, Trattner, K\BPBI J.%
, Toledo-Redondo, S.%
, André, M.%
\BDBL {}Burch, J\BPBI L.%
\end{APACrefauthors}%
\unskip\
\newblock
\APACrefYearMonthDay{2019}{}{}.
\newblock
{\BBOQ}\APACrefatitle {High-density $O^+$ in Earth's outer magnetosphere and
  its effect on dayside magnetopause magnetic reconnection} {High-density $o^+$
  in earth's outer magnetosphere and its effect on dayside magnetopause
  magnetic reconnection}.{\BBCQ}
\newblock
\APACjournalVolNumPages{Journal of Geophysical Research: Space
  Physics}{124}{12}{10257--10269}.
\newblock
\begin{APACrefDOI} \doi{https://doi.org/10.1029/2019JA027396} \end{APACrefDOI}
\PrintBackRefs{\CurrentBib}

\bibitem [\protect \citeauthoryear {%
Fuselier%
\ \protect \BOthers {.}}{%
Fuselier%
\ \protect \BOthers {.}}{%
{\protect \APACyear {2017}}%
}]{%
Fuselier2017}
\APACinsertmetastar {%
Fuselier2017}%
\begin{APACrefauthors}%
Fuselier, S\BPBI A.%
, Vines, S\BPBI K.%
, Burch, J\BPBI L.%
, Petrinec, S\BPBI M.%
, Trattner, K\BPBI J.%
, Cassak, P.%
\BDBL {}Webster, J\BPBI M.%
\end{APACrefauthors}%
\unskip\
\newblock
\APACrefYearMonthDay{2017}{}{}.
\newblock
{\BBOQ}\APACrefatitle {Large-scale characteristics of reconnection diffusion
  regions and associated magnetopause crossings observed by MMS} {Large-scale
  characteristics of reconnection diffusion regions and associated magnetopause
  crossings observed by mms}.{\BBCQ}
\newblock
\APACjournalVolNumPages{Journal of Geophysical Research: Space
  Weather}{122}{}{5466-5486}.
\newblock
\begin{APACrefDOI} \doi{doi:10.1002/2017JA024024.} \end{APACrefDOI}
\PrintBackRefs{\CurrentBib}

\bibitem [\protect \citeauthoryear {%
Ganushkina%
, Liemohn%
\BCBL {}\ \BBA {} Dubyagin%
}{%
Ganushkina%
\ \protect \BOthers {.}}{%
{\protect \APACyear {2018}}%
}]{%
Ganushkina2017}
\APACinsertmetastar {%
Ganushkina2017}%
\begin{APACrefauthors}%
Ganushkina, N\BPBI Y.%
, Liemohn, M\BPBI W.%
\BCBL {}\ \BBA {} Dubyagin, S.%
\end{APACrefauthors}%
\unskip\
\newblock
\APACrefYearMonthDay{2018}{}{}.
\newblock
{\BBOQ}\APACrefatitle {Current Systems in the Earth's Magnetosphere} {Current
  systems in the earth's magnetosphere}.{\BBCQ}
\newblock
\APACjournalVolNumPages{Reviews of Geophysics}{56}{}{309--332}.
\newblock
\begin{APACrefDOI} \doi{https://doi.org/10.1002/2017RG000590} \end{APACrefDOI}
\PrintBackRefs{\CurrentBib}

\bibitem [\protect \citeauthoryear {%
Haaland%
\ \protect \BOthers {.}}{%
Haaland%
\ \protect \BOthers {.}}{%
{\protect \APACyear {2020}}%
}]{%
Haaland2020}
\APACinsertmetastar {%
Haaland2020}%
\begin{APACrefauthors}%
Haaland, S.%
, Paschmann, G.%
, Øieroset, M.%
, Phan, T.%
, Hasegawa, H.%
, Fuselier, S\BPBI A.%
\BDBL {}Burch, J.%
\end{APACrefauthors}%
\unskip\
\newblock
\APACrefYearMonthDay{2020}{}{}.
\newblock
{\BBOQ}\APACrefatitle {Characteristics of the Flank Magnetopause: MMS Results}
  {Characteristics of the flank magnetopause: Mms results}.{\BBCQ}
\newblock
\APACjournalVolNumPages{Journal of Geophysical Research: Space
  Physics}{125}{}{}.
\newblock
\begin{APACrefDOI} \doi{https://doi.org/ 10.1029/2019JA027623} \end{APACrefDOI}
\PrintBackRefs{\CurrentBib}

\bibitem [\protect \citeauthoryear {%
Haaland%
\ \protect \BOthers {.}}{%
Haaland%
\ \protect \BOthers {.}}{%
{\protect \APACyear {2014}}%
}]{%
Haaland2014}
\APACinsertmetastar {%
Haaland2014}%
\begin{APACrefauthors}%
Haaland, S.%
, Reistad, J.%
, Tenfjord, P.%
, Gjerloev, J.%
, Maes, L.%
, DeKeyser, J.%
\BDBL {}Dorville, N.%
\end{APACrefauthors}%
\unskip\
\newblock
\APACrefYearMonthDay{2014}{}{}.
\newblock
{\BBOQ}\APACrefatitle {Characteristics of the flank magnetopause: Cluster
  observations} {Characteristics of the flank magnetopause: Cluster
  observations}.{\BBCQ}
\newblock
\APACjournalVolNumPages{Journal of Geophysical Research: Space
  Physics}{119}{}{9019--9037}.
\newblock
\begin{APACrefDOI} \doi{doi:10.1002/2014JA020539} \end{APACrefDOI}
\PrintBackRefs{\CurrentBib}

\bibitem [\protect \citeauthoryear {%
Haaland%
, Runov%
, Artemyev%
\BCBL {}\ \BBA {} Angelopoulos%
}{%
Haaland%
\ \protect \BOthers {.}}{%
{\protect \APACyear {2019}}%
}]{%
Haaland2019}
\APACinsertmetastar {%
Haaland2019}%
\begin{APACrefauthors}%
Haaland, S.%
, Runov, A.%
, Artemyev, A.%
\BCBL {}\ \BBA {} Angelopoulos, V.%
\end{APACrefauthors}%
\unskip\
\newblock
\APACrefYearMonthDay{2019}{}{}.
\newblock
{\BBOQ}\APACrefatitle {Characteristics of the Flank Magnetopause: THEMIS
  Observations} {Characteristics of the flank magnetopause: Themis
  observations}.{\BBCQ}
\newblock
\APACjournalVolNumPages{Journal of Geophysical Research: Space
  Physics}{124}{}{3421--3435}.
\newblock
\begin{APACrefDOI} \doi{https://doi. org/10.1029/2019JA026459} \end{APACrefDOI}
\PrintBackRefs{\CurrentBib}

\bibitem [\protect \citeauthoryear {%
Hasegawa%
}{%
Hasegawa%
}{%
{\protect \APACyear {2012}}%
}]{%
Hasegawa2012}
\APACinsertmetastar {%
Hasegawa2012}%
\begin{APACrefauthors}%
Hasegawa, H.%
\end{APACrefauthors}%
\unskip\
\newblock
\APACrefYearMonthDay{2012}{}{}.
\newblock
{\BBOQ}\APACrefatitle {Structure and Dynamics of the Magnetopause and Its
  Boundary Layers} {Structure and dynamics of the magnetopause and its boundary
  layers}.{\BBCQ}
\newblock
\APACjournalVolNumPages{Monogr. Environ. Earth Planets}{1}{2}{71-119}.
\newblock
\begin{APACrefDOI} \doi{10.5047/meep.2012.00102.0071} \end{APACrefDOI}
\PrintBackRefs{\CurrentBib}

\bibitem [\protect \citeauthoryear {%
Khrabrov%
\ \BBA {} Sonnerup%
}{%
Khrabrov%
\ \BBA {} Sonnerup%
}{%
{\protect \APACyear {1998a}}%
}]{%
Khrabrov1998a}
\APACinsertmetastar {%
Khrabrov1998a}%
\begin{APACrefauthors}%
Khrabrov, A\BPBI V.%
\BCBT {}\ \BBA {} Sonnerup, B\BPBI U\BPBI O.%
\end{APACrefauthors}%
\unskip\
\newblock
\APACrefYearMonthDay{1998a}{}{}.
\newblock
{\BBOQ}\APACrefatitle {deHoffmann-Teller analysis} {dehoffmann-teller
  analysis}.{\BBCQ}
\newblock
\BIn{} \APACrefbtitle {Analysis methods for multi-spacecraft data} {Analysis
  methods for multi-spacecraft data}\ (\BPG~221).
\newblock
\APACaddressPublisher{Noordwijk}{ESA Publication Division}.
\PrintBackRefs{\CurrentBib}

\bibitem [\protect \citeauthoryear {%
Le%
\ \BBA {} Russell%
}{%
Le%
\ \BBA {} Russell%
}{%
{\protect \APACyear {1994}}%
}]{%
Le1994}
\APACinsertmetastar {%
Le1994}%
\begin{APACrefauthors}%
Le, G.%
\BCBT {}\ \BBA {} Russell, C\BPBI T.%
\end{APACrefauthors}%
\unskip\
\newblock
\APACrefYearMonthDay{1994}{}{}.
\newblock
{\BBOQ}\APACrefatitle {The thickness and structure of high beta magnetopause
  current layer} {The thickness and structure of high beta magnetopause current
  layer}.{\BBCQ}
\newblock
\APACjournalVolNumPages{Geophysical Research Letters}{21}{23}{2451-2454}.
\newblock
\begin{APACrefDOI} \doi{https://doi.org/10.1029/94GL02292} \end{APACrefDOI}
\PrintBackRefs{\CurrentBib}

\bibitem [\protect \citeauthoryear {%
Paschmann%
\ \protect \BOthers {.}}{%
Paschmann%
\ \protect \BOthers {.}}{%
{\protect \APACyear {2018}}%
}]{%
Paschmann2018}
\APACinsertmetastar {%
Paschmann2018}%
\begin{APACrefauthors}%
Paschmann, G.%
, Haaland, S\BPBI E.%
, Phan, T.%
, Sonnerup, B.%
, Burch, J.%
, Torbert, R.%
\BDBL {}Fuselier, S.%
\end{APACrefauthors}%
\unskip\
\newblock
\APACrefYearMonthDay{2018}{}{}.
\newblock
{\BBOQ}\APACrefatitle {Large-Scale Survey of the Structure of the Dayside
  Magnetopause by MMS} {Large-scale survey of the structure of the dayside
  magnetopause by mms}.{\BBCQ}
\newblock
\APACjournalVolNumPages{Journal of Geophysical Research: Space
  Physics}{123}{}{2018--2033}.
\newblock
\begin{APACrefDOI} \doi{https://doi.org/10.1002/2017JA025121} \end{APACrefDOI}
\PrintBackRefs{\CurrentBib}

\bibitem [\protect \citeauthoryear {%
Paschmann%
, Sonnerup%
, Haaland%
, Phan%
\BCBL {}\ \BBA {} Denton%
}{%
Paschmann%
\ \protect \BOthers {.}}{%
{\protect \APACyear {2020}}%
}]{%
Paschmann2020}
\APACinsertmetastar {%
Paschmann2020}%
\begin{APACrefauthors}%
Paschmann, G.%
, Sonnerup, B.%
, Haaland, S\BPBI E.%
, Phan, T.%
\BCBL {}\ \BBA {} Denton, R\BPBI E.%
\end{APACrefauthors}%
\unskip\
\newblock
\APACrefYearMonthDay{2020}{}{}.
\newblock
{\BBOQ}\APACrefatitle {Comparison of Quality Measures for Walén Relation}
  {Comparison of quality measures for walén relation}.{\BBCQ}
\newblock
\APACjournalVolNumPages{Journal of Geophysical Research: Space
  Physics}{125}{e2020JA028044}{}.
\newblock
\begin{APACrefDOI} \doi{https://doi.org/ 10.1029/2020JA028044} \end{APACrefDOI}
\PrintBackRefs{\CurrentBib}

\bibitem [\protect \citeauthoryear {%
Phan%
\ \protect \BOthers {.}}{%
Phan%
\ \protect \BOthers {.}}{%
{\protect \APACyear {1996}}%
}]{%
PhanLarson1996}
\APACinsertmetastar {%
PhanLarson1996}%
\begin{APACrefauthors}%
Phan, T\BPBI D.%
, Larson, D\BPBI E.%
, Lin, R\BPBI P.%
, McFadden, J\BPBI P.%
, Anderson, K\BPBI A.%
, Carlson, C\BPBI W.%
\BDBL {}Szabo, A.%
\end{APACrefauthors}%
\unskip\
\newblock
\APACrefYearMonthDay{1996}{}{}.
\newblock
{\BBOQ}\APACrefatitle {The subsolar magnetosheath and magnetopause for high
  solar wind ram pressure: WIND observations} {The subsolar magnetosheath and
  magnetopause for high solar wind ram pressure: Wind observations}.{\BBCQ}
\newblock
\APACjournalVolNumPages{Geophysical Research Letters}{23}{10}{1279-1282}.
\newblock
\begin{APACrefDOI} \doi{https://doi.org/10.1029/96GL00845} \end{APACrefDOI}
\PrintBackRefs{\CurrentBib}

\bibitem [\protect \citeauthoryear {%
Phan%
\ \BBA {} Paschmann%
}{%
Phan%
\ \BBA {} Paschmann%
}{%
{\protect \APACyear {1996}}%
}]{%
PhanPaschmann1996}
\APACinsertmetastar {%
PhanPaschmann1996}%
\begin{APACrefauthors}%
Phan, T\BPBI D.%
\BCBT {}\ \BBA {} Paschmann, G.%
\end{APACrefauthors}%
\unskip\
\newblock
\APACrefYearMonthDay{1996}{}{}.
\newblock
{\BBOQ}\APACrefatitle {Low-latitude dayside magnetopause and boundary layer for
  high magnetic shear: 1. Structure and motion} {Low-latitude dayside
  magnetopause and boundary layer for high magnetic shear: 1. structure and
  motion}.{\BBCQ}
\newblock
\APACjournalVolNumPages{Journal of Geophysical Research: Space
  Physics}{101}{A4}{7801-7815}.
\newblock
\begin{APACrefDOI} \doi{https://doi.org/10.1029/95JA03752} \end{APACrefDOI}
\PrintBackRefs{\CurrentBib}

\bibitem [\protect \citeauthoryear {%
Pollock%
\ \protect \BOthers {.}}{%
Pollock%
\ \protect \BOthers {.}}{%
{\protect \APACyear {2016}}%
}]{%
FPI}
\APACinsertmetastar {%
FPI}%
\begin{APACrefauthors}%
Pollock, C.%
, Moore, T.%
, Jacques, A.%
, Burch, J.%
, Gliese, U.%
, Saito, Y.%
\BDBL {}et al%
\end{APACrefauthors}%
\unskip\
\newblock
\APACrefYearMonthDay{2016}{}{}.
\newblock
{\BBOQ}\APACrefatitle {Fast Plasma Investigation for Magnetospheric Multiscale}
  {Fast plasma investigation for magnetospheric multiscale}.{\BBCQ}
\newblock
\APACjournalVolNumPages{Space Science Reviews}{199}{}{331-406}.
\newblock
\begin{APACrefDOI} \doi{https://doi.org/10.1007/s11214-016-0245-4}
  \end{APACrefDOI}
\PrintBackRefs{\CurrentBib}

\bibitem [\protect \citeauthoryear {%
Russell%
\ \protect \BOthers {.}}{%
Russell%
\ \protect \BOthers {.}}{%
{\protect \APACyear {2016}}%
}]{%
FGM}
\APACinsertmetastar {%
FGM}%
\begin{APACrefauthors}%
Russell, C\BPBI T.%
, Anderson, B\BPBI J.%
, Baumjohann, W.%
, Bromund, K\BPBI R.%
, Dearborn, D.%
, Fischer, D.%
\BDBL {}Richter, I.%
\end{APACrefauthors}%
\unskip\
\newblock
\APACrefYearMonthDay{2016}{}{}.
\newblock
{\BBOQ}\APACrefatitle {The Magnetospheric Multiscale Magnetometers} {The
  magnetospheric multiscale magnetometers}.{\BBCQ}
\newblock
\APACjournalVolNumPages{Space Science Reviews}{199}{}{189-256}.
\newblock
\begin{APACrefDOI} \doi{https://doi.org/10.1007/s11214-014-0057-3}
  \end{APACrefDOI}
\PrintBackRefs{\CurrentBib}

\bibitem [\protect \citeauthoryear {%
Shuster%
\ \protect \BOthers {.}}{%
Shuster%
\ \protect \BOthers {.}}{%
{\protect \APACyear {2019}}%
}]{%
Shuster2019}
\APACinsertmetastar {%
Shuster2019}%
\begin{APACrefauthors}%
Shuster, J\BPBI R.%
, Gershman, D\BPBI J.%
, Chen, L\BHBI J.%
, Wang, S.%
, Bessho, N.%
, Dorelli, J\BPBI C.%
\BDBL {}Viñas, A\BPBI F.%
\end{APACrefauthors}%
\unskip\
\newblock
\APACrefYearMonthDay{2019}{}{}.
\newblock
{\BBOQ}\APACrefatitle {MMS Measurements of the Vlasov Equation: Probing the
  Electron Pressure Divergence Within Thin Current Sheets} {Mms measurements of
  the vlasov equation: Probing the electron pressure divergence within thin
  current sheets}.{\BBCQ}
\newblock
\APACjournalVolNumPages{Geophysical Research Letters}{46}{}{}.
\newblock
\begin{APACrefDOI} \doi{https://doi.org/10.1029/2019GL083549} \end{APACrefDOI}
\PrintBackRefs{\CurrentBib}

\bibitem [\protect \citeauthoryear {%
Shuster%
\ \protect \BOthers {.}}{%
Shuster%
\ \protect \BOthers {.}}{%
{\protect \APACyear {2021}}%
}]{%
Shuster2021}
\APACinsertmetastar {%
Shuster2021}%
\begin{APACrefauthors}%
Shuster, J\BPBI R.%
, Gershman, D\BPBI J.%
, Dorelli, J\BPBI C.%
, Giles, B\BPBI L.%
, Wang, S.%
, Bessho, N.%
\BDBL {}Torbert, R\BPBI B.%
\end{APACrefauthors}%
\unskip\
\newblock
\APACrefYearMonthDay{2021}{}{}.
\newblock
{\BBOQ}\APACrefatitle {Structures in the terms of the Vlasov equation observed
  at Earth's magnetopause} {Structures in the terms of the vlasov equation
  observed at earth's magnetopause}.{\BBCQ}
\newblock
\APACjournalVolNumPages{Nature Physics}{}{}{}.
\newblock
\begin{APACrefDOI} \doi{https://doi.org/10.1038/s41567-021-01280-6}
  \end{APACrefDOI}
\PrintBackRefs{\CurrentBib}

\bibitem [\protect \citeauthoryear {%
Sonnerup%
\ \BBA {} Scheible%
}{%
Sonnerup%
\ \BBA {} Scheible%
}{%
{\protect \APACyear {1998}}%
}]{%
Sonnerup1998}
\APACinsertmetastar {%
Sonnerup1998}%
\begin{APACrefauthors}%
Sonnerup, B\BPBI U\BPBI O.%
\BCBT {}\ \BBA {} Scheible, M.%
\end{APACrefauthors}%
\unskip\
\newblock
\APACrefYearMonthDay{1998}{}{}.
\newblock
{\BBOQ}\APACrefatitle {Minimum and maximum variance analysis} {Minimum and
  maximum variance analysis}.{\BBCQ}
\newblock
\BIn{} \APACrefbtitle {Analysis methods for multi-spacecraft data} {Analysis
  methods for multi-spacecraft data}\ (\BPG~1850).
\newblock
\APACaddressPublisher{Noordwijk}{ESA Publication Division}.
\PrintBackRefs{\CurrentBib}

\bibitem [\protect \citeauthoryear {%
Walsh%
, Sibeck%
, Nishimura%
\BCBL {}\ \BBA {} Angelopoulos%
}{%
Walsh%
\ \protect \BOthers {.}}{%
{\protect \APACyear {2013}}%
}]{%
Walsh2013}
\APACinsertmetastar {%
Walsh2013}%
\begin{APACrefauthors}%
Walsh, B\BPBI M.%
, Sibeck, D\BPBI G.%
, Nishimura, Y.%
\BCBL {}\ \BBA {} Angelopoulos, V.%
\end{APACrefauthors}%
\unskip\
\newblock
\APACrefYearMonthDay{2013}{}{}.
\newblock
{\BBOQ}\APACrefatitle {Statistical analysis of the plasmaspheric plume at the
  magnetopause} {Statistical analysis of the plasmaspheric plume at the
  magnetopause}.{\BBCQ}
\newblock
\APACjournalVolNumPages{Journal of Geophysical Research: Space
  Physics}{118}{8}{4844-4851}.
\newblock
\begin{APACrefDOI} \doi{https://doi.org/10.1002/jgra.50458} \end{APACrefDOI}
\PrintBackRefs{\CurrentBib}

\bibitem [\protect \citeauthoryear {%
Young%
\ \protect \BOthers {.}}{%
Young%
\ \protect \BOthers {.}}{%
{\protect \APACyear {2016}}%
}]{%
HPCA}
\APACinsertmetastar {%
HPCA}%
\begin{APACrefauthors}%
Young, D\BPBI T.%
, Burch, J\BPBI L.%
, Gomez, R\BPBI G.%
, Santos, A\BPBI D\BPBI L.%
, Miller, G\BPBI P.%
, IV, P\BPBI W.%
\BDBL {}Webster, J\BPBI M.%
\end{APACrefauthors}%
\unskip\
\newblock
\APACrefYearMonthDay{2016}{}{}.
\newblock
{\BBOQ}\APACrefatitle {Hot Plasma Composition Analyzer for the Magnetospheric
  Multiscale Mission} {Hot plasma composition analyzer for the magnetospheric
  multiscale mission}.{\BBCQ}
\newblock
\APACjournalVolNumPages{Space Science Reviews}{199}{}{407-470}.
\newblock
\begin{APACrefDOI} \doi{https://doi.org/10.1007/s11214-014-0119-6}
  \end{APACrefDOI}
\PrintBackRefs{\CurrentBib}

\end{thebibliography}

%
%
%
%
%

\end{document}